\documentclass[useAMS,usenatbib]{mn2e}
\usepackage{graphicx,amsmath,amssymb,times}
\makeatletter

\newcommand{\Rmnum}[1]{\expandafter\@slowromancap\romannumeral #1@}
\makeatother

\title[Long term variability of blazars]{The uncorrelated long term $\gamma$-ray and X-ray variability of blazars and its implications on disk-jet coupling}
\author[Bhattacharya et al.]{Debbijoy Bhattacharya$^{1}$\thanks{E-mail:
debbijoy.b@manipal.edu}, Ranjeev Misra$^{2}$, A.R. Rao$^{3}$, P. Sreekumar$^{4}$\\
$^{1}$Manipal Centre for Natural Sciences, Manipal University, Manipal - 576104, Karnataka, India\\
$^{2}$Inter-University Centre for Astronomy and Astrophysics, Pune - 411007, India \\
$^{3}$Tata Institute of Fundamental Research, Mumbai - 400005, India \\
$^{4}$Space Astronomy Group, ISRO Satellite Centre, Bangalore - 560017, India}

\begin{document}


\pagerange{\pageref{firstpage}--\pageref{lastpage}} \pubyear{2012}

\maketitle

\label{firstpage}

\begin{abstract}
We examine the long term ($\sim$ 10 years) $\gamma$-ray variability of blazars observed by 
EGRET and Fermi and find that  for six sources the average flux  varied by more than an order of 
magnitude. For two of these sources (PKS 0208-512 and PKS 0528+134), there were extensive observations (at various observing periods)  
by  EGRET. Hence these dramatic variations are not due to 
a single short time-scale flare, but reflect long term changes in the average flux. Over the last twenty years, 
these two sources were also the target of several X-ray observatories (e.g. ROSAT, ASCA, RXTE, BeppoSAX, Chandra, 
 Suzaku, XMM-Newton and Swift). 
While the  ratios of the average 
$\gamma$-ray fluxes between EGRET and Fermi observations are $22.9\pm1.9$ and $12.6\pm1.5$, their estimated
$2-10$ keV X-ray flux do not show such dramatic variations.  
 The X-ray emission from such flat spectrum radio 
quasars (FSRQs) are believed to be due to synchrotron self Compton, while $\gamma$-rays 
originate from inverse Comptonization of external soft photons from an accretion 
disk and/or broad line region. We argue that in this scenario, the only explanation for the 
uncorrelated variability is that there was an order of magnitude decrease in the external 
soft photons, while the jet parameters remained more or less constant. This result indicates 
that perhaps the accretion and jet processes are not tightly coupled in these sources.

\end{abstract}

\section{Introduction}
Blazars are a subclass of active galactic nuclei (AGN) whose  bolometric luminosity is dominated by $\gamma$-rays which sometimes extends 
to TeV energies. According to the unification scenario \citep{urry1995}, they are jet dominated sources having a 
small jet to line-of-sight angle. They show 
high variability throughout the electromagnetic spectrum, particularly in  $\gamma$-rays which are believed to be
emitted  close to the base of the jet. However, the nature and emission mechanisms of the jet  are
still  not well understood \citep[e.g.][]{ferrari1998,boettcher2007a,marscher2009}.
$\gamma$-ray observations of these sources can provide a better understanding of fundamental 
issues relating to the energetics, origin of the jet emission and possible connection to the underlying 
accretion disk.

COS-B satellite detected the first extragalactic $\gamma$-ray source (3C 273; \citealt{bignami1981}). 
The first all sky survey in $\gamma$-ray was conducted by EGRET onboard CGRO \citep{kanbach1988,hartman1999}. 
EGRET detected 271 discrete sources above 100 MeV and  $\sim$100 of the identified 
sources of EGRET catalogue are blazars 
(\citealp{hartman1999}; \citealp*{sowards2003}; \citealp{sowards2004}).
Based on their optical line emission properties, blazars can be 
divided into two sub classes: flat spectrum radio quasars (FSRQs) and BL Lacs with the
FSRQs being more luminous \citep[e.g.][]{krolik1999book,kembhavi1999book}. 
EGRET 
detected significantly higher number of FSRQs than BL Lacs. The Fermi $\gamma$-ray space telescope (Fermi), launched in 2008, has $\sim 30$ 
times better sensitivity \citep{atwood2009} and sky coverage than EGRET, and hence it has detected 
$\sim 1900$ sources from its 
more than two years of observations (Second Fermi catalogue: \citealp{Fermi-LATcat2}). 
Most of the identified Fermi sources are AGNs and  the clean sample of 2nd LAT-AGN catalogue 
contains $886$ sources \citep{ackermann2011}. Majority of these AGNs are blazars 
 -- $395$ BL Lacs, $310$ FSRQs, and  $157$ candidate blazars of unknown type.

There have been several attempts to model spectral energy distributions (SEDs) 
of blazars (\citealt{boettcher2004} and the references therein). For example,
modelling the SED of the well known FSRQ 3C 279 \citet{sikora2001}, 
concluded that X-rays and $\gamma$-rays might be produced co-spatially by electrons of similar 
energies. They considered both synchrotron self Compton (SSC) and external Compton (EC) emission 
processes in their model and their results indicated that while X-rays are mainly produced 
via the SSC, $\gamma$-rays are dominated by EC process. This has been confirmed by using
more detailed models and for other FSRQs \citep[e.g.,][]{abdo2010b,palma2011,sahayanathan2012a,mukherjee1999}.

Blazars show variability in $\gamma$-rays over different time scales ranging from flares as short as $\sim$ $1$ hour to 
flares extending for days or months \citep{hartman1999,mattox1997,hartman2001,boettcher2004}. 
Blazars are also variable on still longer timescales of years or decades. The most striking evidence for this is
that a fraction of the EGRET detected blazars are  still not detected by the more sensitive Fermi telescope.
Moreover, some EGRET blazars (e.g., PKS 0208-512) were found to be mostly in the low state during Fermi observations.
While there have been several studies regarding the rapid variability and flaring activity of blazars and their
possible connection to varying jet parameters \citep[e.g.][]{boettcher2003,sahayanathan2012a}, it is not clear whether
the very long term ($\sim 10$ years) variations are similar or if they are intrinsically different phenomena.

In this work we study the long term variability ($\sim 10$ year) of FSRQ that were monitored by 
EGRET during its operation (1991-1997) by comparing their average EGRET flux  with those measured recently by Fermi i.e. after
2007. From 1997 to 2007, there was 
no $\gamma$-ray observatory available to study these sources at GeV energies ($\gamma$-ray satellite AGILE was launched 
in 2007, but no long term monitoring has been reported).There were
pointed X-ray observations of these sources by different satellites (e.g., ROSAT, RXTE-PCA, BeppoSAX, SWIFT, etc.). 
Our motivation is to examine any correlation between the $\gamma$-rays and X-rays, which in turn would provide 
important clues to the main driver of the long term variability. We concentrate on the $\gamma$-ray and X-ray variability,
since SED modelling indicates that although they both arise from the same region (indeed the same electron distribution), their radiative
mechanisms are different and hence their response to variation in the intrinsic jet properties will be different.

\citet{abdo2010c} and \citet{palma2011} studied the relatively short term 
variability of blazars utilising quasi-simultaneous multiwaveband observations. 
\citet{abdo2010c} carried out quasi simultaneous multiwaveband observations of $5$ 
$\gamma$-ray loud blazars between 2008 October and 2009 January. Using the result from SED modelling they argue that 
the difference between the low and high state of luminous blazars is due to the different kinetic power of the jet which 
is most likely related to the varying bulk Lorentz factor of outflow within the blazar emission zone.  
\citet{palma2011} reported that PKS 0528+134 did not show any significant variability 
in $\gamma$-ray band on $1-2$ weeks timescales. They found moderate variability in X-rays ($ |\frac{\Delta F}{F}| \sim 50\%$) 
and most radio frequencies ($ |\frac{\Delta F}{F}| \lesssim 20\% $) on timescales of $1-2$ weeks. In optical bands the 
variability is found to be of upto $ \Delta R \lesssim 1$ on timescale of several hours. 
However, the physics driving the long term variability may be 
different than the short term one.

\section{Observations}

From the 3rd  EGRET point source catalogue
\citep{hartman1999}, blazars have been identified using multi-wavelength information
(\citealp{hartman1999}; \citealp*{mattox2001}). Using a different identification technique
\citet{sowards2003,sowards2004} have obtained a more complete census 
of plausible blazar counterparts. We use their source list augmented with source identifications from 3rd EGRET catalogue 
in regions of the sky that were not covered by \citet{sowards2003,sowards2004}. 
We further consider only FSRQs that have  a cumulative 
source detection significance 
$> 4 \sigma$ for sources above the Galactic plane ($|\theta| > 10^{\circ}$) and  $> 5 \sigma$ for those in the Galactic plane 
($-10^{\circ} \le \theta \le 10^{\circ}$ ).
The $\gamma$-ray fluxes averaged over all the cycles of EGRET observations (P1234) were used for our calculations. The P1234 
flux has been calculated from the counts map over four cycles.
 Our final source list contains $56$ FSRQs out of which $38$ 
are also present in the 2nd Fermi catalogue. To be compatible with the Fermi data,
we consider only EGRET $\gamma$-ray fluxes above $1$ GeV, which are calculated
from the average spectral index and $> 100$ MeV flux quoted in the 3rd EGRET 
catalogue. 

We find $5$ FSRQs which are brighter by an order of magnitude or
more during EGRET era compared to the average flux observed from
Fermi data while there is one source that is  fainter by a factor of
$\sim 10$ (Table 1).

\begin{table}
\begin{center}
\caption{Details of highly variable  blazars\label{tbl-1}}
\begin{tabular}{lccc}
\hline
Source Name & F$_{\mbox{EGRET}}$$^{a}$ & F$_{\mbox{Fermi}}$$^{a}$ & $R$$^{a}$\\
\hline
PKS 0208-512 & $86.7\pm 4.6$ & $3.79\pm 0.24$ & $22.9\pm 1.9$ \\
3EGJ 1614+3424 & $10.1\pm 1.5$ & $0.49\pm 0.11$ & $20.6\pm 5.5$ \\
PKS 1622-29 &$40.1\pm 3.1$ & $2.40\pm 0.26$ & $16.7\pm 2.2$ \\
PKS 0336-01 &$21.4\pm 5.0$ & $1.35\pm 0.17$ &$15.9\pm 4.2$ \\
PKS 0528+134 & $32.4\pm 1.2$ & $2.57\pm 0.29$ &$12.6\pm 1.5$ \\
PKS 1454-354 & $1.1\pm 0.3$ &$11.9\pm 0.42$ & $0.09\pm 0.02$ \\
\hline
\end{tabular}
\end{center}
$^{a}${$\gamma$-ray ($> 1$ GeV) flux during EGRET and Fermi observations in $10^{-9}$ photon cm$^{-2}$ s$^{-1}$.$R$ is
the ratio of the EGRET to Fermi flux.}

\end{table}

\begin{figure}

\includegraphics[angle=0,height=3cm,width=8cm]{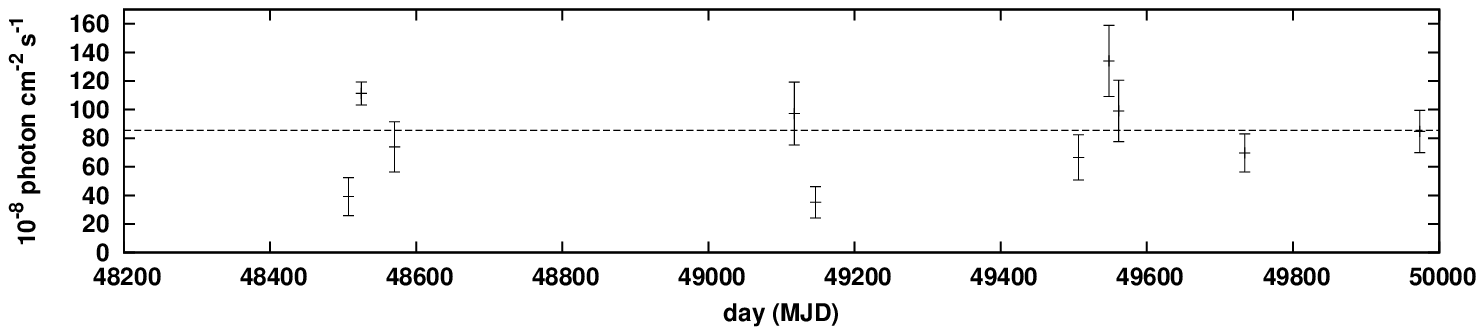}
\includegraphics[angle=0,height=3cm,width=8cm]{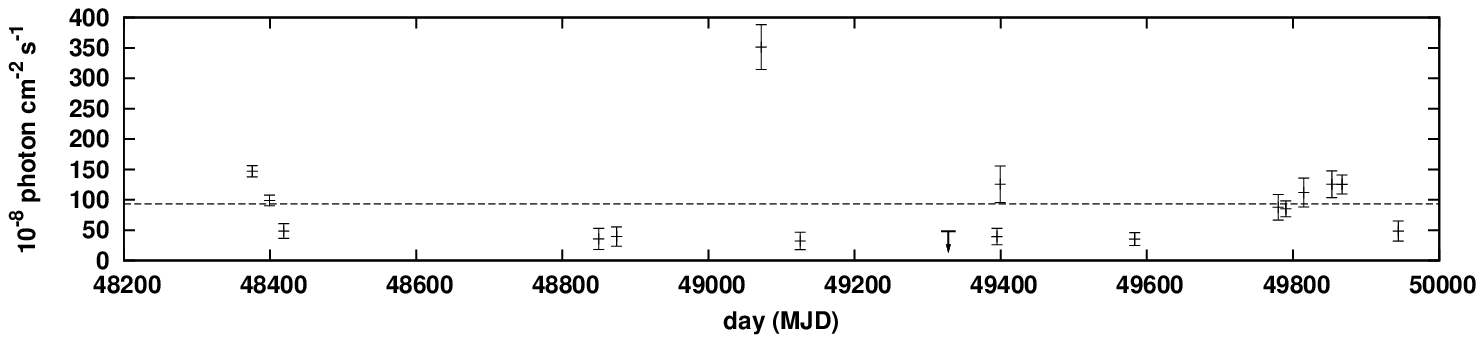}
\caption{EGRET lightcurve ($> 100$ MeV) of PKS 0208-512 (top panel) and  PKS 0528+134 (bottom panel) averaged over each viewing period. The dash line indicates the average P1234 flux value. 
MJD 48200 and MJD 50000 correspond to 1990 November 5 and 1995 October 10 respectively. \label{egretlc}}
\end{figure}

 We focus first on two of these
highly variable six sources (PKS 0208-512 and PKS 0528+134) that have
been  detected in several EGRET viewing cycles. These two sources were
also observed in X-rays by different satellites during EGRET to
Fermi era. Figure \ref{egretlc} top and bottom panels show
EGRET lightcurves of PKS 0208-512 and PKS 0528+134 respectively. The
dash line indicates  the P1234 average flux value. The figure shows that
there was extensive coverage of these sources by EGRET and the average values
are not dominated by a single, short time-scale flare.

\begin{table*}
\begin{center}
\caption{Summary of X-ray observations of PKS 0208-512 and PKS 0528+134 \label{xobs1}}
\begin{tabular}{cccccc}
\hline
Satellite & Date & Energy band (keV) & Flux$^{a}$ & Flux$_{2-10 keV}$$^{b}$ & Ref.$^{c}$    \\
\hline
PKS 0208-512 & && &   \\
ROSAT & 1990-1991 & $0.1-2.4$  & $1.72 \pm 0.36$ & $0.98 \pm 0.21$ &1  \\
 ASCA & 1995 & $0.5-10$ & $9.49$ & $6.76$                &2 \\
 ASCA & 1996 & $2-10$ & $5.77$ & $5.77$                 & 3 \\
 BeppoSax & 2001 & $2-10$ & $4.7$ & $4.7$               & 4,5 \\
 Chandra & 2001 & $2-10$  &$2.1^{d} $ & $2.1^{d}$                 & 6  \\
 SWIFT & 2005 & $0.5-10$ & $3.4$ & $1.9$               & 7,8 \\
 SWIFT & 2008 & $0.5-10$ & $2.7$ & $1.6$                 & 7,8 \\
Suzaku & 2008 & $2-10$ &$1.37 \pm 0.06$ & $1.37 \pm 0.06$ & 7 \\~\\
PKS 0528+134 & && &   \\
  ROSAT & 1990-1991 & $0.1-2.4$ & $3.13\pm0.63$ & $3.99 \pm 0.80$ &  1,9,10\\
  ASCA & 1994 & $2-10$ & $3.5^{+0.8}_{-0.4}$ & $3.5^{+0.8}_{-0.4}$ &  11\\
  ASCA & 1995 & $2-10$ & $8.98$ & $8.98$ & 3\\
 RXTE-PCA & 1996 & $2-10$ & $10.62\pm0.70$$^{e}$ & $10.62\pm 0.70$$^{ e}$ & -\\
  BeppoSax & 1997 & $2-10$ & $2.5$ & $2.5$ &  5,12\\
  RXTE-PCA & 1999 & $2-10$ & $7.00\pm 0.58$$^{e}$ & $7.00\pm0.58$$^{e}$ & -\\
 Suzaku&2008 & $2-10$ & $2.8$ & $2.8$                              & 13  \\
RXTE-PCA & 2009  & $2-10$ & $< 7.85$$^{e}$ & $< 7.85$$^{e}$         &   -  \\
 XMM-Newton & 2009 & $0.2-2$ & $0.28^{+0.03}_{-0.04}$ & $1.35\pm0.11$ & 13 \\
 SWIFT & 2009 & $0.2-10$ & $4.0 \pm 0.6$ & $3.0 \pm 0.5$            & 14 \\
\hline

\end{tabular}
\end{center}
{$^{a}$Reported X-ray flux  in $10^{-12}$ erg cm$^{-2}$s$^{-1}$ in the energy band given in column 3. 
$^{b}$Using the fluxes and spectral parameters 
observed in different energy bands given in references the X-ray fluxes are converted in 2-10 keV energy band in $10^{-12}$ erg cm$^{-2}$s$^{-1}$ unit. 
$^{c}$ References: 1: \citet*{brinkmann1994} 2: \citet{reeves2000} 3: \citet{ueda2001} 4: \citet{tavecchio2002} 5: \citet*{donato2005}. 6: \citet{schwartz2006} 
7: \citet{abdo2010c} 8: \citet{zhang2010} 9: \citet{zhang1994} 10: \citet{mukherjee1996} 
11: \citet{sambruna1997} 12: \citet{ghisellini1999} 13: \citet{palma2011} 14: \citet*{ghisellini2009b} 
$^{d}$Flux has been calculated from the rest-frame $2-10$ keV luminosity given in the reference. 
$^{e}$Averaged over the observations shown in Fig \ref{rxtelc1}

}

\end{table*}

 \begin{figure}
\includegraphics[angle=0,height=3cm,width=8cm]{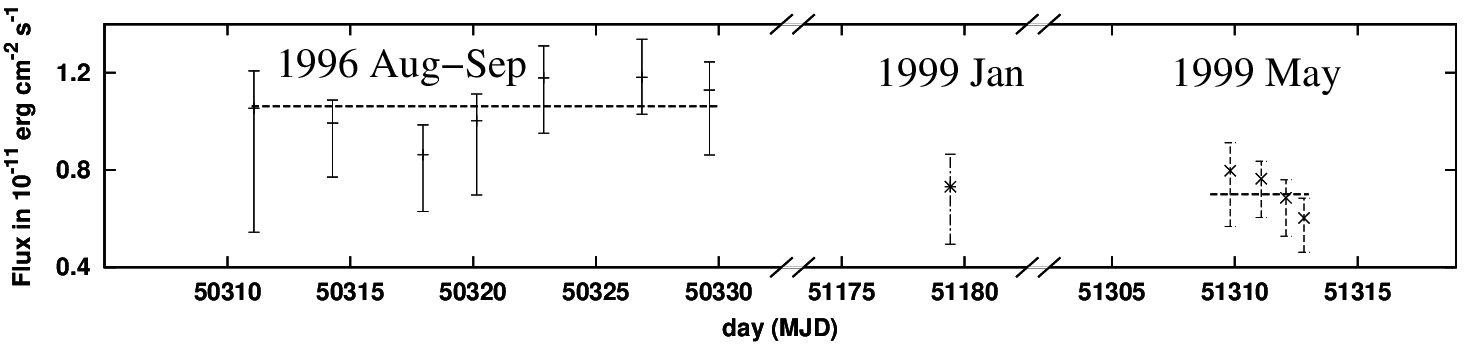}
\includegraphics[angle=0,height=3cm,width=8cm]{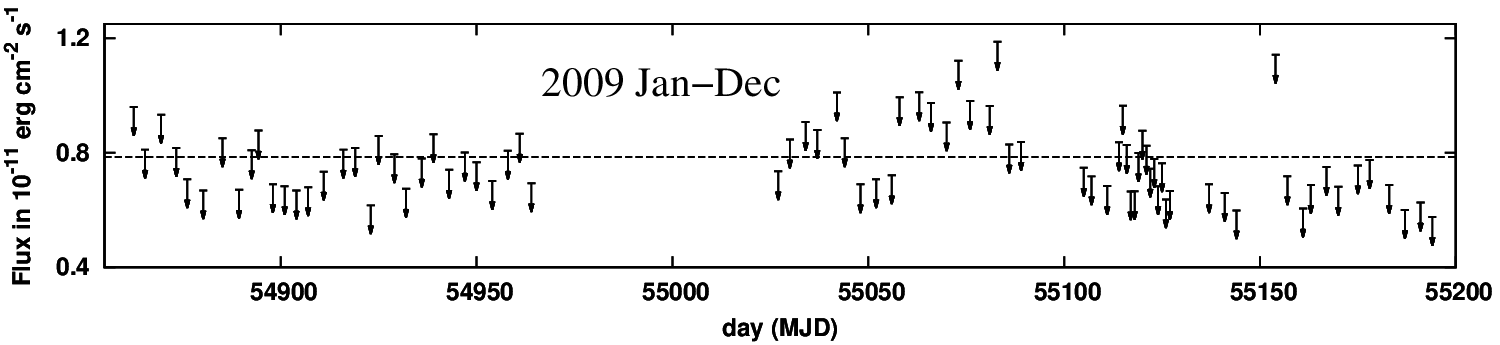}
\caption {RXTE-PCA lightcurve of PKS 0528+134. The left-hand branch of points of the top panel represents the observations during 
1996 August and September, and the middle and right-hand branch represent observations during 1999 January and May, respectively. 
The bottom panel represents observations during 2009. The horizontal dash line indicates the average flux over the subset of data.\label{rxtelc1}}
\end{figure}

\begin{figure}

\includegraphics[angle=0,height=6cm,width=9cm]{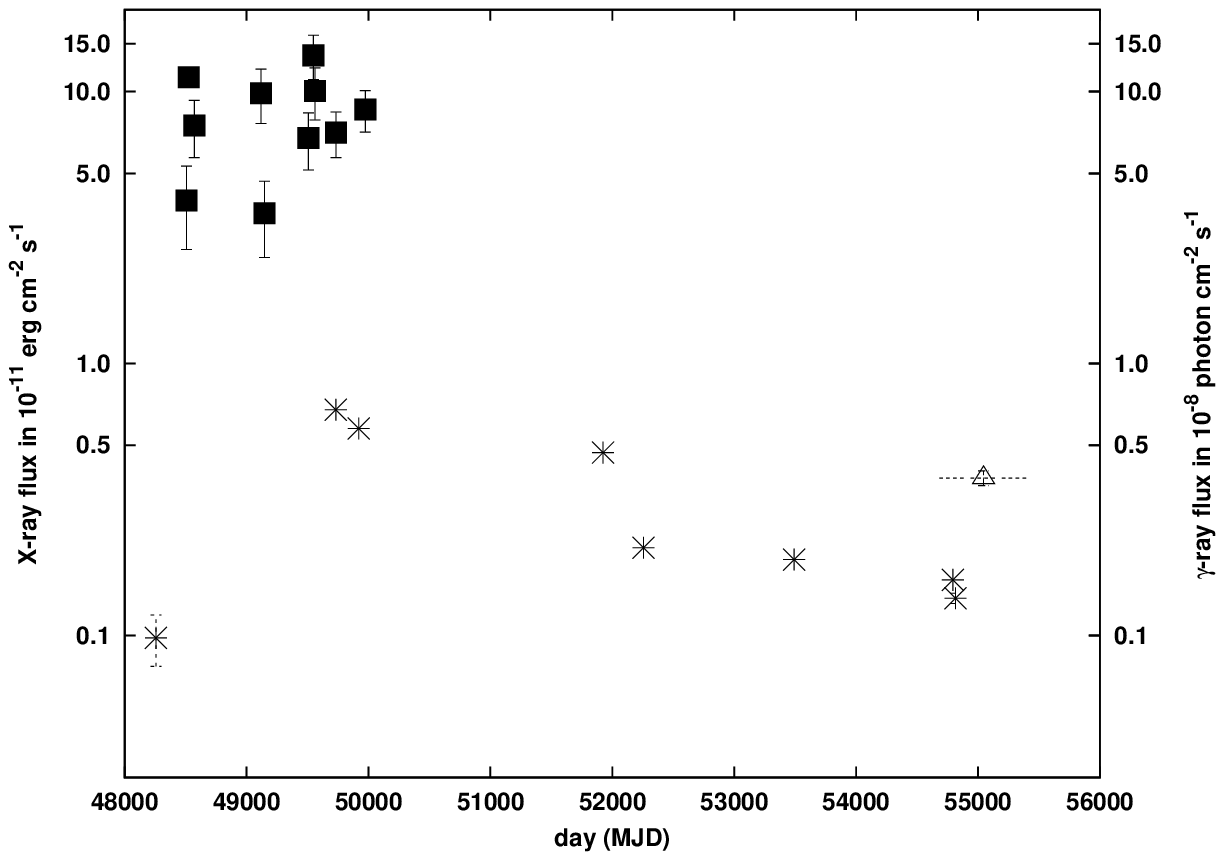}
\includegraphics[angle=0,height=6cm,width=9cm]{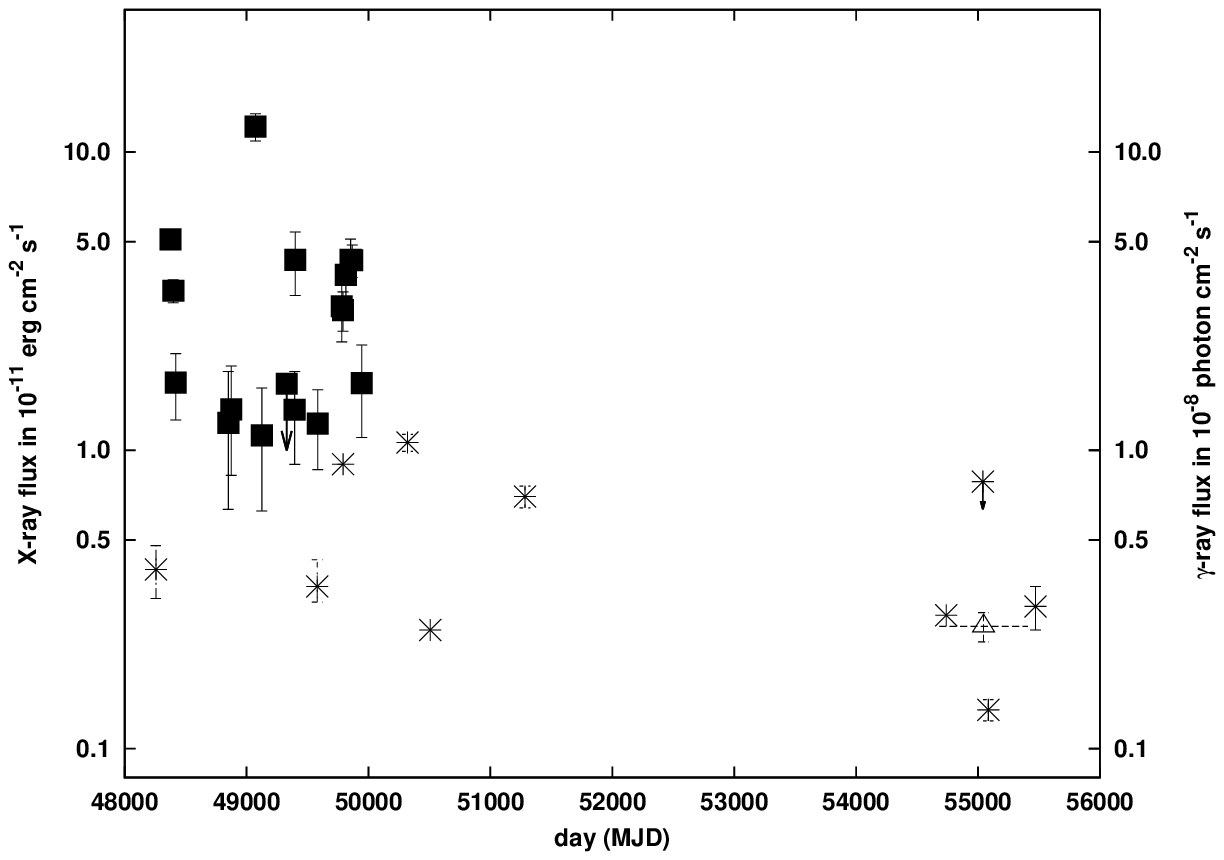}
\caption{
X-ray and $\gamma$-ray fluxes of PKS 0208-512 (top panel) and  PKS 0528+134 (bottom panel) as observed by different satellites. 
The stars represent X-ray observations from different instruments from EGRET to Fermi era. RXTE fluxes are averaged over 
observations shown in Fig \ref{rxtelc1}.  It is evident that 
average X-ray flux did not decrease by more than a factor of $2$ from EGRET to Fermi era. 
The triangle represents the $\gamma$-ray flux ($>1$ GeV) averaged over first two year of Fermi observations. The boxes represent 
EGRET fluxes ($>1$ GeV) averaged over each observing period ($\sim 2$ weeks). The scale for gamma-ray fluxes are given in the
right side of the figure. 
MJD 48000 and MJD 56000 correspond to 1990 April 19 and 2012 March 14 respectively. \label{S1}}
\end{figure}

PKS 0528+134 is one of the most luminous  blazars detected by 
EGRET and showed several flaring episodes from 1990 to 1997 \citep{mukherjee1997a}.  It was also detected by both OSSE 
and COMPTEL onboard CGRO \citep{mcnaron1995,collmar1997} and has been observed 
simultaneously at other wavelengths (\citealp{mukherjee1999} and references therein).
PKS 0208-512 shows the maximum flux variations with an average EGRET flux that is $22.9\pm1.9$ times 
higher than the average Fermi flux. 
Over the last twenty years, both sources  were observed by several X-ray satellites and  Table \ref{xobs1} gives their X-ray fluxes as reported. 
These different instruments measured X-ray fluxes in different energy bands (column 3 of Table \ref{xobs1}).
Using the reported spectral parameters, we converted the observed flux in different energy bands into flux in the 2-10 keV band. Specifically, we use the
reported column density and spectral index to model the spectrum in XSPEC and extracted the 2-10 keV flux (column 5 of Table \ref{xobs1}).

Pointed observation by RXTE Proportional counter array (PCA) of PKS 0528+134 were undertaken 
in 1996, 1999 and 2009. 
We used the spectra available in the standard product of the PCA observations, modelled with a power-law and Galactic absorption
to obtain a $2-10$ keV integrated flux. Figure \ref{rxtelc1} shows the flux variations across different observations. During 2009, 
only upper limits were found probably because only one of the Proportional Counter Unit  was working during this time. During
these observations the flux did not vary significantly and hence we report an average value for three periods of observations
in Table \ref{xobs1}. 

Figures \ref{S1} compares the $\gamma$-ray and X-ray fluxes at different times for the two sources. 
EGRET $\gamma$-ray fluxes are averaged over each observing period ($\sim 2$ weeks) and the Fermi flux 
is averaged over the first two years of 
observation. X-ray fluxes are from pointed observations. While there is
considerable scatter in the X-ray fluxes, there is no evidence for a dramatic decrease in X-ray flux, especially
for PKS 0528+134. For PKS 0208-512, there is a hint of a decrease in the X-ray flux if one neglects the early ROSAT
point. However, this decrease is at most a factor of 3, which is significantly less than the variation in the $\gamma$-ray flux
of a factor of  23. Thus, we conclude that the large variability observed in $\gamma$-rays is not seen in X-rays.

\begin{table*}
\begin{center}
\caption{Summary of X-ray observations of other four FSRQs \label{xobs2}}
\begin{tabular}{cccccc}
\hline
 Satellite & Date & Energy band (keV) & Flux$^{a}$ & Flux$_{2-10 keV}$$^{b}$ & Ref.$^{c}$    \\
\hline
PKS 1622-29 & && &   \\
ROSAT & 1990-1991 & $0.1-2.4$ & $0.32 \pm 0.08$ & $0.31 \pm 0.08$     & 1,2 \\
 RXTE-PCA & 1997 & $2-10$ & $11.95 \pm 0.26$ & $11.95\pm0.26$ &  - \\
RXTE-PCA & 2006 & $2-10$ & $14.49 \pm 0.25$ & $14.49\pm0.25$ &  - \\~\\
PKS 0336-01 & && &   \\
Einstein & 1979-1980 & $0.15-3.5$ & $0.48 \pm 0.20$ & $0.68 \pm 0.28$ & 3  \\
 ROSAT & 1990-1991 & $0.1-2.4$ & $0.98 \pm 0.29$ & $0.73 \pm 0.22$ &  1  \\~\\
PKS 1454-354 & && &   \\
 ROSAT & 1990-91 & $0.1-2.4$ &  $0.510$ & $0.290$ & 1,4 \\
 SWIFT & 2008 January & $2-10$ & $0.6\pm0.12$ & $0.6\pm0.12$ & 5 \\
 SWIFT & 2008 September & $2-10$ & $1.8\pm0.4$ & $1.8\pm0.4$ & 5 \\~\\
3EGJ 1614+3424 & && &   \\
ROSAT  & 1990-91 & $0.1-2.4$  & $0.401$ & $0.228$ & 1,4 \\
\hline
\end{tabular}
\end{center}
{$^{a}$Reported X-ray flux  in $10^{-12}$ erg cm$^{-2}$s$^{-1}$ in the energy band given in column 3.
$^{b}$Using fluxes and spectral parameters 
observed in different energy bands given in references X-ray fluxes are derived for a common 2-10 keV energy band in $10^{-12}$ erg cm$^{-2}$s$^{-1}$ unit. 
$^{c}$ References: 1: \citet{brinkmann1994} 2: \citet{mattox1997} 3: \citet*{owen1981} 4: \citet{massaro2009} 5: \citet{ghisellini2009b}
}
\end{table*}

For the other four sources with high $\gamma$-ray 
flux variability, only sparse X-ray observations are available. Nevertheless we report all available X-ray fluxes in Table \ref{xobs2}.
For PKS 1622-29, we have analysed existing RXTE PCA observations of 1997 and 2000. As compared
to the early ROSAT observation, X-ray flux has increased in contrast to $\gamma$-ray emission which decreased. However,
the small number of observations do not permit a reliable conclusion. For the rest, there are no reported X-ray observations
during both EGRET and Fermi era. 

Statistical errors on the measured flux have been reported for some of the observations. However, due to the
uncertainties in the relative normalisations between different instruments and in the extrapolation to the
2-10 keV band, systematic errors on flux could be higher, but not expected to be larger than a factor
of 2.

For two of the blazars, PKS 0528+134 and PKS 1622-29, the RXTE All Sky Monitor (ASM) lightcurves are available in the
RXTE website (http://xte.mit.edu/asmlc/One-Day.html). However, the long term time averaged count rates for these sources PKS 0528+134 and PKS 1622-29 
are not significantly higher than
the systematic positive bias expected in ASM data  \citep{chitnis2009}. Thus ASM did not detect these sources.

 It is to be noted that X-ray coverage of these sources are quite sparse and hence, the actual short term variability
 can be under-sampled. 
However, our main aim is to search for any long term 
variability which is less affected by the under-sampling 
of light curves. Also, for PKS 0528+134, quasi-simultaneous gamma-ray 
and X-ray observations have been carried out during almost all EGRET 
observations. It was observed that apart from a single huge flare, 
the variation in both X-ray and gamma-ray emission during these times 
are less than a factor of $3$. 

\section{Discussion}
In leptonic jet models, non-thermal electrons in the 
jet produce synchrotron emission. These electrons  upscatter 
the synchrotron photons (SSC) and also  photons coming out 
from accretion disk and/or broad line region (BLR) (EC) via the inverse Compton process. 
These emission processes contribute to the SED of blazars
and  modelling suggests that $\gamma$-ray emissions of FSRQs are dominated by 
EC process  while the X-rays are primarily due to SSC 
(e.g., 3C 279: \citealp{pian1999,abdo2010b,sahayanathan2012a}, PKS 0528+134: \citealp{palma2011}).
On the other hand, SEDs of BL Lac sources are better fitted by 
synchrotron self Compton (SSC) emission (e.g., Mrk 501: \citealp{petry2000,abdo2010b}). Recent progress in theoretical models 
for blazar emission is  summarised by \citet{boettcher2010a}. 

Specifically, \citet{mukherjee1999} carried out a multiwaveband study of PKS 0528+134 
during EGRET observations and  SED modelling showed that during the $\gamma$-ray high state, 
the $\gamma$-ray emission was due to 
 external Compton (EC) emission. 
Using Fermi observations, \citet{palma2011} also carried out a coordinated multi-wavelength campaign 
and similarly concluded that as expected for FSRQs, the $\gamma$-ray emission is mainly due to EC process 
while SSC process is the primary X-ray emission mechanism.

In this scenario, we investigate variability in which jet or environment parameters can give rise
to a strong variation in the EC component (i.e. $\gamma$-rays)  while not affecting the SSC one (i.e. X-rays).
Variation of the magnetic field density is ruled out since that would affect SSC and not the EC emission. 
Similarly, a change in the number density of electron $n$ would affect SSC more ($\propto n^2$) than the
EC emission ($\propto n$).  Since, blazar emission is highly beamed, 
observed flux is a sensitive function of the Doppler beaming factor ($\delta$). 
For a discrete jet, synchrotron self Compton emission is proportional to $\delta^{(3+ \alpha)}$, 
whereas, external Compton flux is proportional to $\delta^{(4+ 2\alpha)}$, where  $\alpha$ is the 
energy spectral index. The average value of $\gamma$-ray photon spectral index ($\Gamma = \alpha +1$) derived from Fermi and EGRET data sets 
are $2.47 \pm 0.19$ and $2.34 \pm 0.15$ respectively (\citealp{abdo2010e}; \citealp*{deb2009a}) and hence variation in $\alpha$ cannot
be the cause of the $\gamma$-ray intensity variation. For the average value of $\alpha = 1.4$, the Doppler factor $\delta$ 
would need to decrease by $\sim 1.4$ to cause a factor of $10$ decrease in the EC flux. However, this would
have led to a decrease in the SSC component as well by a factor of $\sim 5$, which was not observed. Thus,
variation in any of the intrinsic jet parameters such as number density, magnetic field, spectral index or
Doppler factor cannot explain the uncorrelated variation in X-ray/$\gamma$-ray emission.

The EC emission is proportional to the external photon  flux from accretion disk or BLR region, while SSC emission 
is independent. If the external photon flux was a factor of $\sim 10$ more during EGRET observation time, then it would 
naturally explain why $\gamma$-rays changed by a factor of ten while the X-ray flux remained nearly
constant.  This has important implications on the relationship between the jet and the accretion process.
If the flux from the disk or BLR, which is connected to the accretion process can change by an order of
magnitude but the average intrinsic properties of the jet are not affected, it implies very weak or no 
coupling between the accretion and jet emission processes. 

For PKS 0528+134, the  X-ray observations showed no significant variation i.e. the variation was less than a factor of $2$. 
If one considers the extreme conservative limit, wherein the X-ray emission did
decrease by a factor of $2$ during the Fermi era, it implies that the Doppler factor did decrease by a
factor $\sim 1.17$. However, this decrease in Doppler factor alone cannot explain the observed large $\gamma$-ray flux variation. 
An additional decrease in the external photon flux by a factor of  $\sim 5$ is required.
Thus in this extreme case, a factor of $5$ change in the accretion process caused only a modest $\sim 1.17$ factor change 
in the Doppler factor, implying that the accretion and jet processes are not tightly coupled.

If the external photons arising from the accretion disk or the BLR region did indeed change significantly over time, one can observe a 
significant decrease in the optical broad line flux during the Fermi era compared to EGRET. Such a decrease, would
be a direct evidence for the model proposed here.   
However, in order to obtain an averaged line flux long term spectroscopic monitoring of these sources in optical is required. 
While, there were several photometric observations of PKS 0528+134  \citep[e.g.][]{villata1997,raiteri1998,mukherjee1999}, 
optical spectroscopic observations have been  carried out only in recent years \citep{palma2011}. 

 Apart from synchrotron process, the photometric observation will have contributions from accretion disk, 
BLR region and also from the host galaxies. So 
the optical photometric flux is not always dominated by synchrotron and/or synchrotron 
self Compton processes (\citealp*{whiting2003}; \citealp{mukherjee1999}). 
PKS 0528+134 was observed in optical during few EGRET observing periods and the 
variation is within a factor of $\sim 3$.  This source was also observed in optical during 
the night of 2001 August 29 \citep{whiting2003}. \citet{palma2011} 
also observed the source in 2009 and the variation between these 
observations (BVR band) is a factor of $\sim 2$. 

There were only few optical observations of PKS 0208-512. This source was observed 
in optical wavebands during the nights of 2001 August 29 \citep{whiting2003} and 
2008 December 14 \citep{abdo2010c}. So, it is not possible to have a meaningful 
long term average of optical emission from this object.

PKS 1622-29 was monitored in optical (mainly in R band) for more than a decade by 
the program for Extragalactic Astronomy (PEGA) \citep{meyer2008}. 
Though the source showed strong variation in shorter time scale (Figure 1 and 2 of 
\citealp{meyer2008}), the long term average variation is within a factor of $\sim 3$. 
However, the number of pointed observations is too few to derive a reliable long term
average flux. More importantly, a huge $\gamma$-ray flare of this source was observed with EGRET 
and hence, the calculated EGRET average flux may be overestimated. 
Therefore, we have mainly considered 
PKS 0208-512 and PKS 0528+134 for our analysis. Unfortunately, 
as per our knowledge, there are no reported optical spectra of these blazars during both  EGRET and Fermi times.

Radio emission probably originates in a different emission 
region than that of X-ray and $\gamma$-ray. Hence, the high frequency 
radio flux might be delayed by a few months to that of X-rays and $\gamma$-rays. 
Since the main objective of this work is to study the long term behaviour 
of blazars ($\sim 10$ year), due to the longer cooling time scales at 
radio wavelengths, the high-frequency radio power could be a good 
representative for the long time-averaged jet power. 
PKS 0528+134 has been observed at a wavelength of $850 \mu m$ 
from 1997 April to 2004 December using SCUBA camera at 
the James Clerk Maxwell Telescope \citep*{jenness2010}. 
A total of $129$ observations were carried out 
over $62$ nights. The source did not show large variations during this period 
($\frac{Flux_{max}}{Flux_{min}} < 3$). \citet*{lonsdale1998} 
conducted a survey of bright 
extragalactic radio sources with $3$ mm ($100$ GHz) VLBI. 
They reported that the $3$ mm fluxes are $2.9$ Jy and 
$2.5$ Jy as measured by Kitt Peak and Haystack telescopes 
and MPIR $100$ meter telescope near Bonn and Onsala telescopes 
respectively. This source has been studied in the observed 
frequency range $80-267$ GHz from 2007 February to 2008 December 
with IRAM Plateau de Bure Interferometer (PdBI). Reported 
average fluxes in $1.3$ mm band, $2$ mm band and $3$ mm band are 
($2\pm1.1$) Jy, ($1.7\pm0.5$) Jy and ($3.6\pm1.3$) Jy respectively \citep{maury2010}. 
Clearly, the flux variation in $3$ mm band between these two 
observations is less than a factor of $2$. \citet{agudo2010} 
observed PKS 0528+134 during 2005 July at $86$ GHz and 
the reported flux is ($2.0\pm0.10$) Jy. This source was also 
monitored at $86$ GHz during 2001 October and the reported 
flux is $2.02$ Jy \citep{lee2008}. This source is 
also observed at other high radio frequencies at different 
times and no large flux variation is noticed 
(e.g., \citealp{palma2011}; \citealp{massardi2009}; \citealp*{gu2009}; \citealp{lanyi2010}; 
\citealp{wiren1992}; \citealp{abdo2010b}; \citealp{petrov2007}; 
\citealp{lister2005}; \citealp{kellermann2004}; \citealp{cara2008} etc.). 

The measured $150$ GHz flux of PKS 0208-512 is $1268\pm86$ mJy 
utilising the observation with Arcminute Cosmology Bolometer 
Array Receiver (ACBAR) during 2002 observing seasons 
\citep{kuo2007}. WMAP has detected PKS 0208-512 and no 
significant variation is noticed among one year averaged, 
three year averaged and five year averaged fluxes 
\citep{bennett2003,hinshaw2007,wright2009}. Using the five year WMAP data, 
\citet{chen2009} reported that observed $41$ GHz, $61$ GHz 
and $94$ GHz fluxes of this source are ($2.7\pm0.2$) Jy, 
($2.7\pm0.2$) Jy and ($1.8\pm0.4$) Jy respectively. 
PKS 0208-512 has been observed at other high radio frequencies 
at different times and no large flux variation is noticed 
(e.g., \citealp{ricci2006}; \citealp{massardi2008}; 
\citealp{murphy2010}; \citealp{tingay2003};  etc.). 

The results obtained in this work need to be confirmed for other highly variable $\gamma$-ray sources through
reliable long term X-ray monitoring data. It would also be interesting to study long term behaviour of 
such sources at other wavelengths which may provide additional clues on the strength of coupling between the jet and accretion process.

\section*{Acknowledgements}
This work is based on the data provided by the ASM/RXTE
teams at MIT and at the RXTE SOF, GOF and High Energy Astrophysics Science Archive Research
Center (HEASARC) provided by NASA’s
Goddard Space Flight Center. 
\def\apj{ApJ}%
\def\mnras{MNRAS}%
\def\aap{A\&A}%
\def\apjl{ApJ}
\def\aj{AJ}
\def\physrep{PhR}
\def\apjs{ApJS}
\def\pasa{PASA}
\def\pasj{PASJ}
\def\nat{Natur}
\def\apss{Ap\&SS}
\def\araa{ARA\&A}
\def\aaps{A\&AS}
\def\ssr{Space Sci. Rev.}
\def\pasp{PASP}
\bibliographystyle{mn2emod}
\bibliography{debbijoy_blazar.bib}

\begin{thebibliography}{74}
\expandafter\ifx\csname natexlab\endcsname\relax\def\natexlab#1{#1}\fi

\bibitem[{{Abdo} {et~al}\mbox{.}(2010{\natexlab{a}}){Abdo}, {Ackermann},
  {Agudo}, {Ajello}, {Aller}, {Aller}, {Angelakis}, {Arkharov}, {Axelsson},
  {Bach}, \& et~al.}]{abdo2010b}
{Abdo} A.~A. {et~al.}, 2010{\natexlab{a}}, \apj, 716, 30

\bibitem[{{Abdo} {et~al}\mbox{.}(2010{\natexlab{b}}){Abdo}, {Ackermann},
  {Ajello}, {Antolini}, {Baldini}, {Ballet}, {Barbiellini}, {Baring},
  {Bastieri}, {Bechtol}, {Bellazzini}, {Berenji}, {Blandford}, {Bloom},
  {Bonamente}, {Borgland}, {Bregeon}, {Brez}, {Brigida}, {Bruel}, {Buehler},
  {Buson}, {Caliandro}, {Cameron}, {Carrigan}, {Casandjian}, {Cavazzuti},
  {Cecchi}, {{\c C}elik}, {Chekhtman}, {Chen}, {Chiang}, {Ciprini}, {Claus},
  {Cohen-Tanugi}, {Colafrancesco}, {Conrad}, {Cutini}, {Dermer}, {de Palma},
  {Digel}, {Silva}, {Drell}, {Dubois}, {Dumora}, {Farnier}, {Favuzzi}, {Fegan},
  {Ferrara}, {Focke}, {Frailis}, {Fukazawa}, {Fusco}, {Gargano}, {Gasparrini},
  {Gehrels}, {Giebels}, {Giglietto}, {Giommi}, {Giordano}, {Giroletti},
  {Glanzman}, {Godfrey}, {Grandi}, {Grenier}, {Guillemot}, {Guiriec},
  {Hadasch}, {Harding}, {Hayashida}, {Horan}, {Hughes}, {Itoh}, {Jackson},
  {J{\'o}hannesson}, {Johnson}, {Johnson}, {Kamae}, {Katagiri}, {Kataoka},
  {Kawai}, {Kn{\"o}dlseder}, {Kuss}, {Lande}, {Latronico}, {Longo}, {Loparco},
  {Lott}, {Lovellette}, {Lubrano}, {Madejski}, {Makeev}, {Mazziotta},
  {McEnery}, {McGlynn}, {Meurer}, {Michelson}, {Mitthumsiri}, {Mizuno},
  {Monte}, {Monzani}, {Morselli}, {Moskalenko}, {Murgia}, {Nestoras}, {Nolan},
  {Norris}, {Nuss}, {Ohsugi}, {Okumura}, {Orlando}, {Ormes}, {Ozaki},
  {Paneque}, {Panetta}, {Parent}, {Pelassa}, {Pepe}, {Pesce-Rollins}, {Piron},
  {Porter}, {Rain{\`o}}, {Rando}, {Razzano}, {Reimer}, {Reimer}, {Reyes},
  {Rodriguez}, {Roth}, {Ryde}, {Sadrozinski}, {Sambruna}, {Sander}, {Sato},
  {Sgr{\`o}}, {Shaw}, {Siskind}, {Smith}, {Spandre}, {Spinelli}, {Stawarz},
  {Stecker}, {Strickman}, {Suson}, {Takahashi}, {Takahashi}, {Tanaka},
  {Thayer}, {Thayer}, {Thompson}, {Tibolla}, {Torres}, {Tosti}, {Tramacere},
  {Uchiyama}, {Usher}, {Vasileiou}, {Vilchez}, {Villata}, {Vitale}, {von
  Kienlin}, {Waite}, {Wang}, {Winer}, {Wood}, {Yang}, {Ylinen}, {Ziegler},
  {Tavecchio}, {Sikora}, {Schady}, {Roming}, {Chester}, \&
  {Maraschi}}]{abdo2010c}
{Abdo} A.~A. {et~al.}, 2010{\natexlab{b}}, \apj, 716, 835

\bibitem[{{Abdo} {et~al}\mbox{.}(2010{\natexlab{c}}){Abdo}, {Ackermann},
  {Ajello}, {Antolini}, {Baldini}, {Ballet}, {Barbiellini}, {Bastieri},
  {Baughman}, {Bechtol}, {Bellazzini}, {Berenji}, {Blandford}, {Bloom},
  {Bonamente}, {Borgland}, {Bouvier}, {Bregeon}, {Brez}, {Brigida}, {Bruel},
  {Burnett}, {Buson}, {Caliandro}, {Cameron}, {Caraveo}, {Carrigan},
  {Casandjian}, {Cavazzuti}, {Cecchi}, {{\c C}elik}, {Charles}, {Chekhtman},
  {Cheung}, {Chiang}, {Ciprini}, {Claus}, {Cohen-Tanugi}, {Conrad},
  {Costamante}, {Cutini}, {Dermer}, {de Angelis}, {de Palma}, {Silva}, {Drell},
  {Dubois}, {Dumora}, {Farnier}, {Favuzzi}, {Fegan}, {Focke}, {Fukazawa},
  {Funk}, {Fusco}, {Gargano}, {Gasparrini}, {Gehrels}, {Germani}, {Giglietto},
  {Giommi}, {Giordano}, {Glanzman}, {Godfrey}, {Grenier}, {Grove}, {Guiriec},
  {Hadasch}, {Hayashida}, {Hays}, {Healey}, {Horan}, {Hughes}, {Itoh},
  {J{\'o}hannesson}, {Johnson}, {Johnson}, {Johnson}, {Kamae}, {Katagiri},
  {Kataoka}, {Kawai}, {Kn{\"o}dlseder}, {Kuss}, {Lande}, {Latronico}, {Lee},
  {Lemoine-Goumard}, {Llena Garde}, {Longo}, {Loparco}, {Lott}, {Lovellette},
  {Lubrano}, {Madejski}, {Makeev}, {Mazziotta}, {McConville}, {McEnery},
  {Meurer}, {Michelson}, {Mitthumsiri}, {Mizuno}, {Monte}, {Monzani},
  {Morselli}, {Moskalenko}, {Murgia}, {Nolan}, {Norris}, {Nuss}, {Ohsugi},
  {Omodei}, {Orlando}, {Ormes}, {Ozaki}, {Paneque}, {Panetta}, {Parent},
  {Pelassa}, {Pepe}, {Pesce-Rollins}, {Piron}, {Porter}, {Rain{\`o}}, {Rando},
  {Razzano}, {Reimer}, {Reimer}, {Ritz}, {Rochester}, {Rodriguez}, {Romani},
  {Roth}, {Sadrozinski}, {Sander}, {Saz Parkinson}, {Scargle}, {Sgr{\`o}},
  {Shaw}, {Smith}, {Spandre}, {Spinelli}, {Starck}, {Strickman}, {Strong},
  {Suson}, {Tajima}, {Takahashi}, {Takahashi}, {Tanaka}, {Thayer}, {Thayer},
  {Thompson}, {Tibaldo}, {Torres}, {Tosti}, {Tramacere}, {Uchiyama}, {Usher},
  {Vasileiou}, {Vilchez}, {Vitale}, {Waite}, {Wang}, {Winer}, {Wood}, {Yang},
  {Ylinen}, {Ziegler}, \& {Fermi-LAT Collaboration}}]{abdo2010e}
{Abdo} A.~A. {et~al.}, 2010{\natexlab{c}}, \apj, 720, 435

\bibitem[{{Ackermann} {et~al}\mbox{.}(2011){Ackermann}, {Ajello}, {Allafort},
  {Antolini}, {Atwood}, {Axelsson}, {Baldini}, {Ballet}, {Barbiellini},
  {Bastieri}, {Bechtol}, {Bellazzini}, {Berenji}, {Blandford}, {Bloom},
  {Bonamente}, {Borgland}, {Bottacini}, {Bouvier}, {Bregeon}, {Brigida},
  {Bruel}, {Buehler}, {Burnett}, {Buson}, {Caliandro}, {Cameron}, {Caraveo},
  {Casandjian}, {Cavazzuti}, {Cecchi}, {Charles}, {Cheung}, {Chiang},
  {Ciprini}, {Claus}, {Cohen-Tanugi}, {Conrad}, {Costamante}, {Cutini}, {de
  Angelis}, {de Palma}, {Dermer}, {Digel}, {Silva}, {Drell}, {Dubois},
  {Escande}, {Favuzzi}, {Fegan}, {Ferrara}, {Finke}, {Focke}, {Fortin},
  {Frailis}, {Fukazawa}, {Funk}, {Fusco}, {Gargano}, {Gasparrini}, {Gehrels},
  {Germani}, {Giebels}, {Giglietto}, {Giommi}, {Giordano}, {Giroletti},
  {Glanzman}, {Godfrey}, {Grenier}, {Grove}, {Guiriec}, {Gustafsson},
  {Hadasch}, {Hayashida}, {Hays}, {Healey}, {Horan}, {Hou}, {Hughes},
  {Iafrate}, {J{\'o}hannesson}, {Johnson}, {Johnson}, {Kamae}, {Katagiri},
  {Kataoka}, {Kn{\"o}dlseder}, {Kuss}, {Lande}, {Larsson}, {Latronico},
  {Longo}, {Loparco}, {Lott}, {Lovellette}, {Lubrano}, {Madejski}, {Mazziotta},
  {McConville}, {McEnery}, {Michelson}, {Mitthumsiri}, {Mizuno}, {Moiseev},
  {Monte}, {Monzani}, {Moretti}, {Morselli}, {Moskalenko}, {Murgia},
  {Nakamori}, {Naumann-Godo}, {Nolan}, {Norris}, {Nuss}, {Ohno}, {Ohsugi},
  {Okumura}, {Omodei}, {Orienti}, {Orlando}, {Ormes}, {Ozaki}, {Paneque},
  {Parent}, {Pesce-Rollins}, {Pierbattista}, {Piranomonte}, {Piron}, {Pivato},
  {Porter}, {Rain{\`o}}, {Rando}, {Razzano}, {Razzaque}, {Reimer}, {Reimer},
  {Ritz}, {Rochester}, {Romani}, {Roth}, {Sanchez}, {Sbarra}, {Scargle},
  {Schalk}, {Sgr{\`o}}, {Shaw}, {Siskind}, {Spandre}, {Spinelli}, {Strong},
  {Suson}, {Tajima}, {Takahashi}, {Takahashi}, {Tanaka}, {Thayer}, {Thayer},
  {Thompson}, {Tibaldo}, {Tinivella}, {Torres}, {Tosti}, {Troja}, {Uchiyama},
  {Vandenbroucke}, {Vasileiou}, {Vianello}, {Vitale}, {Waite}, {Wallace},
  {Wang}, {Winer}, {Wood}, {Wood}, \& {Zimmer}}]{ackermann2011}
{Ackermann} M. {et~al.}, 2011, \apj, 743, 171

\bibitem[{{Agudo} {et~al}\mbox{.}(2010){Agudo}, {Thum}, {Wiesemeyer}, \&
  {Krichbaum}}]{agudo2010}
{Agudo} I., {Thum} C., {Wiesemeyer} H., {Krichbaum} T.~P., 2010, \apjs, 189, 1

\bibitem[{{Atwood} {et~al}\mbox{.}(2009){Atwood}, {Abdo}, {Ackermann},
  {Althouse}, {Anderson}, {Axelsson}, {Baldini}, {Ballet}, {Band},
  {Barbiellini}, \& et~al.}]{atwood2009}
{Atwood} W.~B. {et~al.}, 2009, \apj, 697, 1071

\bibitem[{{Bennett} {et~al}\mbox{.}(2003){Bennett}, {Hill}, {Hinshaw}, {Nolta},
  {Odegard}, {Page}, {Spergel}, {Weiland}, {Wright}, {Halpern}, {Jarosik},
  {Kogut}, {Limon}, {Meyer}, {Tucker}, \& {Wollack}}]{bennett2003}
{Bennett} C.~L. {et~al.}, 2003, \apjs, 148, 97

\bibitem[{{Bhattacharya} {et~al}\mbox{.}(2009){Bhattacharya}, {Sreekumar}, \&
  {Mukherjee}}]{deb2009a}
{Bhattacharya} D., {Sreekumar} P., {Mukherjee} R., 2009, Research in Astronomy
  and Astrophysics, 9, 85

\bibitem[{{Bignami} {et~al}\mbox{.}(1981){Bignami}, {Bennett}, {Buccheri},
  {Caraveo}, {Hermsen}, {Kanbach}, {Lichti}, {Masnou}, {Mayer-Hasselwander},
  {Paul}, {Sacco}, {Scarsi}, {Swanenburg}, \& {Wills}}]{bignami1981}
{Bignami} G.~F. {et~al.}, 1981, \aap, 93, 71

\bibitem[{{Boettcher}(2004)}]{boettcher2004}
{Boettcher} M., 2004, in ESA Special Publication, Vol. 552, Proceedings of the
  5th INTEGRAL Workshop on the INTEGRAL Universe, 16-20 February 2004, Munich,
  Germany., {V.~Schoenfelder, G.~Lichti, \& C.~Winkler}, ed., p. 543

\bibitem[{{Boettcher}(2010)}]{boettcher2010a}
{Boettcher} M., 2010, ArXiv e-prints, astro-ph:1006.5048

\bibitem[{{B{\"o}ttcher}(2007)}]{boettcher2007a}
{B{\"o}ttcher} M., 2007, \apss, 309, 95

\bibitem[{{B{\"o}ttcher} {et~al}\mbox{.}(2003){B{\"o}ttcher}, {Marscher},
  {Ravasio}, {Villata}, {Raiteri}, {Aller}, {Aller}, {Ter{\"a}sranta}, {Mang},
  {Tagliaferri}, {Aharonian}, {Krawczynski}, {Kurtanidze}, {Nikolashvili},
  {Ibrahimov}, {Papadakis}, {Tsinganos}, {Sadakane}, {Okada}, {Takalo},
  {Sillanp{\"a}{\"a}}, {Tosti}, {Ciprini}, {Frasca}, {Marilli}, {Robb},
  {Noble}, {Jorstad}, {Hagen-Thorn}, {Larionov}, {Nesci}, {Maesano},
  {Schwartz}, {Basler}, {Gorham}, {Iwamatsu}, {Kato}, {Pullen},
  {Ben{\'{\i}}tez}, {de Diego}, {Moilanen}, {Oksanen}, {Rodriguez}, {Sadun},
  {Kelly}, {Carini}, {Miller}, {Catalano}, {Dultzin-Hacyan}, {Fan},
  {Ghisellini}, {Ishioka}, {Karttunen}, {Kein{\"a}nen}, {Kudryavtseva},
  {Lainela}, {Lanteri}, {Larionova}, {Matsumoto}, {Mattox}, {McHardy},
  {Montagni}, {Nucciarelli}, {Ostorero}, {Papamastorakis}, {Pasanen},
  {Sobrito}, \& {Uemura}}]{boettcher2003}
{B{\"o}ttcher} M. {et~al.}, 2003, \apj, 596, 847

\bibitem[{{Brinkmann} {et~al}\mbox{.}(1994){Brinkmann}, {Siebert}, \&
  {Boller}}]{brinkmann1994}
{Brinkmann} W., {Siebert} J., {Boller} T., 1994, \aap, 281, 355

\bibitem[{{Cara} \& {Lister}(2008)}]{cara2008}
{Cara} M., {Lister} M.~L., 2008, \apj, 674, 111

\bibitem[{{Chen} \& {Wright}(2009)}]{chen2009}
{Chen} X., {Wright} E.~L., 2009, \apj, 694, 222

\bibitem[{{Chitnis} {et~al}\mbox{.}(2009){Chitnis}, {Pendharkar}, {Bose},
  {Agrawal}, {Rao}, \& {Misra}}]{chitnis2009}
{Chitnis} V.~R., {Pendharkar} J.~K., {Bose} D., {Agrawal} V.~K., {Rao} A.~R.,
  {Misra} R., 2009, \apj, 698, 1207

\bibitem[{{Collmar} {et~al}\mbox{.}(1997){Collmar}, {Bennett}, {Bloemen},
  {Blom}, {Hermsen}, {Lichti}, {Pohl}, {Ryan}, {Schonfelder}, {Stacy},
  {Steinle}, \& {Williams}}]{collmar1997}
{Collmar} W. {et~al.}, 1997, \aap, 328, 33

\bibitem[{{Donato} {et~al}\mbox{.}(2005){Donato}, {Sambruna}, \&
  {Gliozzi}}]{donato2005}
{Donato} D., {Sambruna} R.~M., {Gliozzi} M., 2005, \aap, 433, 1163

\bibitem[{{Ferrari}(1998)}]{ferrari1998}
{Ferrari} A., 1998, \araa, 36, 539

\bibitem[{{Ghisellini} {et~al}\mbox{.}(1999){Ghisellini}, {Costamante},
  {Tagliaferri}, {Maraschi}, {Celotti}, {Fossati}, {Pian}, {Comastri}, {de
  Francesco}, {Lanteri}, {Raiteri}, {Sobrito}, {Villata}, {Glass}, {Grandi},
  {Perola}, \& {Treves}}]{ghisellini1999}
{Ghisellini} G. {et~al.}, 1999, \aap, 348, 63

\bibitem[{{Ghisellini} {et~al}\mbox{.}(2009){Ghisellini}, {Tavecchio}, \&
  {Ghirlanda}}]{ghisellini2009b}
{Ghisellini} G., {Tavecchio} F., {Ghirlanda} G., 2009, \mnras, 399, 2041

\bibitem[{{Gu} {et~al}\mbox{.}(2009){Gu}, {Cao}, \& {Jiang}}]{gu2009}
{Gu} M., {Cao} X., {Jiang} D.~R., 2009, \mnras, 396, 984

\bibitem[{{Hartman} {et~al}\mbox{.}(1999){Hartman}, {Bertsch}, {Bloom}, {Chen},
  {Deines-Jones}, {Esposito}, {Fichtel}, {Friedlander}, {Hunter}, {McDonald},
  {Sreekumar}, {Thompson}, {Jones}, {Lin}, {Michelson}, {Nolan}, {Tompkins},
  {Kanbach}, {Mayer-Hasselwander}, {M{\"u}cke}, {Pohl}, {Reimer}, {Kniffen},
  {Schneid}, {von Montigny}, {Mukherjee}, \& {Dingus}}]{hartman1999}
{Hartman} R.~C. {et~al.}, 1999, \apjs, 123, 79

\bibitem[{{Hartman} {et~al}\mbox{.}(2001){Hartman}, {Villata}, {Balonek},
  {Bertsch}, {Bock}, {B{\"o}ttcher}, {Carini}, {Collmar}, {De Francesco},
  {Ferrara}, {Heidt}, {Kanbach}, {Katajainen}, {Koskimies}, {Kurtanidze},
  {Lanteri}, {Lawson}, {Lin}, {Marscher}, {McFarland}, {McHardy}, {Miller},
  {Nikolashvili}, {Nilsson}, {Noble}, {Nucciarelli}, {Ostorero}, {Pursimo},
  {Raiteri}, {Rekola}, {Savolainen}, {Sillanp{\"a}{\"a}}, {Smale}, {Sobrito},
  {Takalo}, {Thompson}, {Tosti}, {Wagner}, \& {Wilson}}]{hartman2001}
{Hartman} R.~C. {et~al.}, 2001, \apj, 558, 583

\bibitem[{{Hinshaw} {et~al}\mbox{.}(2007){Hinshaw}, {Nolta}, {Bennett}, {Bean},
  {Dor{\'e}}, {Greason}, {Halpern}, {Hill}, {Jarosik}, {Kogut}, {Komatsu},
  {Limon}, {Odegard}, {Meyer}, {Page}, {Peiris}, {Spergel}, {Tucker}, {Verde},
  {Weiland}, {Wollack}, \& {Wright}}]{hinshaw2007}
{Hinshaw} G. {et~al.}, 2007, \apjs, 170, 288

\bibitem[{{Jenness} {et~al}\mbox{.}(2010){Jenness}, {Robson}, \&
  {Stevens}}]{jenness2010}
{Jenness} T., {Robson} E.~I., {Stevens} J.~A., 2010, \mnras, 401, 1240

\bibitem[{{Kanbach} {et~al}\mbox{.}(1988){Kanbach}, {Bertsch}, {Fichtel},
  {Hartman}, {Hunter}, {Kniffen}, {Hughlock}, {Favale}, {Hofstadter}, \&
  {Hughes}}]{kanbach1988}
{Kanbach} G. {et~al.}, 1988, \ssr, 49, 69

\bibitem[{{Kellermann} {et~al}\mbox{.}(2004){Kellermann}, {Lister}, {Homan},
  {Vermeulen}, {Cohen}, {Ros}, {Kadler}, {Zensus}, \&
  {Kovalev}}]{kellermann2004}
{Kellermann} K.~I. {et~al.}, 2004, \apj, 609, 539

\bibitem[{{Kembhavi}(1999)}]{kembhavi1999book}
{Kembhavi} A., 1999, {Quasar and Active Galactic Nuclei}. Cambridge University
  Press, |1999

\bibitem[{{Krolik}(1999)}]{krolik1999book}
{Krolik} J., 1999, {Active Galactic Nuclei}. Princeton University Press, |1999

\bibitem[{{Kuo} {et~al}\mbox{.}(2007){Kuo}, {Ade}, {Bock}, {Bond}, {Contaldi},
  {Daub}, {Goldstein}, {Holzapfel}, {Lange}, {Lueker}, {Newcomb}, {Peterson},
  {Reichardt}, {Ruhl}, {Runyan}, \& {Staniszweski}}]{kuo2007}
{Kuo} C.~L. {et~al.}, 2007, \apj, 664, 687

\bibitem[{{Lanyi} {et~al}\mbox{.}(2010){Lanyi}, {Boboltz}, {Charlot}, {Fey},
  {Fomalont}, {Geldzahler}, {Gordon}, {Jacobs}, {Ma}, {Naudet}, {Romney},
  {Sovers}, \& {Zhang}}]{lanyi2010}
{Lanyi} G.~E. {et~al.}, 2010, \aj, 139, 1695

\bibitem[{{Lee} {et~al}\mbox{.}(2008){Lee}, {Lobanov}, {Krichbaum}, {Witzel},
  {Zensus}, {Bremer}, {Greve}, \& {Grewing}}]{lee2008}
{Lee} S.-S., {Lobanov} A.~P., {Krichbaum} T.~P., {Witzel} A., {Zensus} A.,
  {Bremer} M., {Greve} A., {Grewing} M., 2008, \aj, 136, 159

\bibitem[{{Lister} \& {Homan}(2005)}]{lister2005}
{Lister} M.~L., {Homan} D.~C., 2005, \aj, 130, 1389

\bibitem[{{Lonsdale} {et~al}\mbox{.}(1998){Lonsdale}, {Doeleman}, \&
  {Phillips}}]{lonsdale1998}
{Lonsdale} C.~J., {Doeleman} S.~S., {Phillips} R.~B., 1998, \aj, 116, 8

\bibitem[{{Marscher}(2009)}]{marscher2009}
{Marscher} A.~P., 2009, ArXiv e-prints, astro-ph: 0909.2576

\bibitem[{{Massardi} {et~al}\mbox{.}(2008){Massardi}, {Ekers}, {Murphy},
  {Ricci}, {Sadler}, {Burke}, {de Zotti}, {Edwards}, {Hancock}, {Jackson},
  {Kesteven}, {Mahony}, {Phillips}, {Staveley-Smith}, {Subrahmanyan}, {Walker},
  \& {Wilson}}]{massardi2008}
{Massardi} M. {et~al.}, 2008, \mnras, 384, 775

\bibitem[{{Massardi} {et~al}\mbox{.}(2009){Massardi}, {L{\'o}pez-Caniego},
  {Gonz{\'a}lez-Nuevo}, {Herranz}, {de Zotti}, \& {Sanz}}]{massardi2009}
{Massardi} M., {L{\'o}pez-Caniego} M., {Gonz{\'a}lez-Nuevo} J., {Herranz} D.,
  {de Zotti} G., {Sanz} J.~L., 2009, \mnras, 392, 733

\bibitem[{{Massaro} {et~al}\mbox{.}(2009){Massaro}, {Giommi}, {Leto},
  {Marchegiani}, {Maselli}, {Perri}, {Piranomonte}, \& {Sclavi}}]{massaro2009}
{Massaro} E., {Giommi} P., {Leto} C., {Marchegiani} P., {Maselli} A., {Perri}
  M., {Piranomonte} S., {Sclavi} S., 2009, \aap, 495, 691

\bibitem[{{Mattox} {et~al}\mbox{.}(2001){Mattox}, {Hartman}, \&
  {Reimer}}]{mattox2001}
{Mattox} J.~R., {Hartman} R.~C., {Reimer} O., 2001, \apjs, 135, 155

\bibitem[{{Mattox} {et~al}\mbox{.}(1997){Mattox}, {Wagner}, {Malkan},
  {McGlynn}, {Schachter}, {Grove}, {Johnson}, \& {Kurfess}}]{mattox1997}
{Mattox} J.~R., {Wagner} S.~J., {Malkan} M., {McGlynn} T.~A., {Schachter}
  J.~F., {Grove} J.~E., {Johnson} W.~N., {Kurfess} J.~D., 1997, \apj, 476, 692

\bibitem[{{Maury} {et~al}\mbox{.}(2010){Maury}, {Andr{\'e}}, {Hennebelle},
  {Motte}, {Stamatellos}, {Bate}, {Belloche}, {Duch{\^e}ne}, \&
  {Whitworth}}]{maury2010}
{Maury} A.~J. {et~al.}, 2010, \aap, 512, A40

\bibitem[{{McNaron-Brown} {et~al}\mbox{.}(1995){McNaron-Brown}, {Johnson},
  {Jung}, {Kinzer}, {Kurfess}, {Strickman}, {Dermer}, {Grabelsky}, {Purcell},
  {Ulmer}, {Kafatos}, {Becker}, {Staubert}, \& {Maisack}}]{mcnaron1995}
{McNaron-Brown} K. {et~al.}, 1995, \apj, 451, 575

\bibitem[{{Mukherjee} {et~al}\mbox{.}(1997){Mukherjee}, {Bertsch}, {Bloom},
  {Dingus}, {Esposito}, {Fichtel}, {Hartman}, {Hunter}, {Kanbach}, {Kniffen},
  {Lin}, {Mayer-Hasselwander}, {McDonald}, {Michelson}, {von Montigny},
  {Muecke}, {Nolan}, {Pohl}, {Reimer}, {Schneid}, {Sreekumar}, \&
  {Thompson}}]{mukherjee1997a}
{Mukherjee} R. {et~al.}, 1997, \apj, 490, 116

\bibitem[{{Mukherjee} {et~al}\mbox{.}(1999){Mukherjee}, {B{\"o}ttcher},
  {Hartman}, {Sreekumar}, {Thompson}, {Mahoney}, {Pursimo},
  {Sillanp{\"a}{\"a}}, \& {Takalo}}]{mukherjee1999}
{Mukherjee} R. {et~al.}, 1999, \apj, 527, 132

\bibitem[{{Mukherjee} {et~al}\mbox{.}(1996){Mukherjee}, {Dingus}, {Gear},
  {Hartman}, {Hunter}, {Marscher}, {Moore}, {Pohl}, {Robson}, {Sreekumar},
  {Stevens}, {Teraesranta}, {Tornikoski}, {Travis}, {Wagner}, \&
  {Zhang}}]{mukherjee1996}
{Mukherjee} R. {et~al.}, 1996, \apj, 470, 831

\bibitem[{{Murphy} {et~al}\mbox{.}(2010){Murphy}, {Sadler}, {Ekers},
  {Massardi}, {Hancock}, {Mahony}, {Ricci}, {Burke-Spolaor}, {Calabretta},
  {Chhetri}, {de Zotti}, {Edwards}, {Ekers}, {Jackson}, {Kesteven}, {Lindley},
  {Newton-McGee}, {Phillips}, {Roberts}, {Sault}, {Staveley-Smith},
  {Subrahmanyan}, {Walker}, \& {Wilson}}]{murphy2010}
{Murphy} T. {et~al.}, 2010, \mnras, 402, 2403

\bibitem[{{Nolan} {et~al}\mbox{.}(2012){Nolan}, {Abdo}, {Ackermann}, {Ajello},
  {Allafort}, {Antolini}, {Atwood}, {Axelsson}, {Baldini}, {Ballet}, \&
  et~al.}]{Fermi-LATcat2}
{Nolan} P.~L. {et~al.}, 2012, \apjs, 199, 31

\bibitem[{{Osterman Meyer} {et~al}\mbox{.}(2008){Osterman Meyer}, {Miller},
  {Marshall}, {Ryle}, {Aller}, {Aller}, {McFarland}, {Pollock}, {Reichart},
  {Crain}, {Ivarsen}, {La Cluyze}, \& {Nysewander}}]{meyer2008}
{Osterman Meyer} A. {et~al.}, 2008, \aj, 136, 1398

\bibitem[{{Owen} {et~al}\mbox{.}(1981){Owen}, {Helfand}, \&
  {Spangler}}]{owen1981}
{Owen} F.~N., {Helfand} D.~J., {Spangler} S.~R., 1981, \apjl, 250, L55

\bibitem[{{Palma} {et~al}\mbox{.}(2011){Palma}, {B{\"o}ttcher}, {de la Calle},
  {Agudo}, {Aller}, {Aller}, {Bach}, {Ben{\'{\i}}tez}, {Buemi}, {Escande},
  {G{\'o}mez}, {Gurwell}, {Heidt}, {Hiriart}, {Jorstad}, {Joshi},
  {L{\"a}hteenm{\"a}ki}, {Larionov}, {Leto}, {Li}, {L{\'o}pez}, {Lott},
  {Madejski}, {Marscher}, {Morozova}, {Raiteri}, {Roberts}, {Tornikoski},
  {Trigilio}, {Umana}, {Villata}, \& {Wylezalek}}]{palma2011}
{Palma} N.~I. {et~al.}, 2011, \apj, 735, 60

\bibitem[{{Petrov} {et~al}\mbox{.}(2007){Petrov}, {Hirota}, {Honma}, {Shibata},
  {Jike}, \& {Kobayashi}}]{petrov2007}
{Petrov} L., {Hirota} T., {Honma} M., {Shibata} K.~M., {Jike} T., {Kobayashi}
  H., 2007, \aj, 133, 2487

\bibitem[{{Petry} {et~al}\mbox{.}(2000){Petry}, {B{\"o}ttcher}, {Connaughton},
  {Lahteenmaki}, {Pursimo}, {Raiteri}, {Schr{\"o}der}, {Sillanp{\"a}{\"a}},
  {Sobrito}, {Takalo}, {Ter{\"a}sranta}, {Tosti}, \& {Villata}}]{petry2000}
{Petry} D. {et~al.}, 2000, \apj, 536, 742

\bibitem[{{Pian} {et~al}\mbox{.}(1999){Pian}, {Urry}, {Maraschi}, {Madejski},
  {McHardy}, {Koratkar}, {Treves}, {Chiappetti}, {Grandi}, {Hartman}, {Kubo},
  {Leach}, {Pesce}, {Imhoff}, {Thompson}, \& {Wehrle}}]{pian1999}
{Pian} E. {et~al.}, 1999, \apj, 521, 112

\bibitem[{{Raiteri} {et~al}\mbox{.}(1998){Raiteri}, {Ghisellini}, {Villata},
  {de Francesco}, {Lanteri}, {Chiaberge}, {Peila}, \& {Antico}}]{raiteri1998}
{Raiteri} C.~M., {Ghisellini} G., {Villata} M., {de Francesco} G., {Lanteri}
  L., {Chiaberge} M., {Peila} A., {Antico} G., 1998, \aaps, 127, 445

\bibitem[{{Reeves} \& {Turner}(2000)}]{reeves2000}
{Reeves} J.~N., {Turner} M.~J.~L., 2000, \mnras, 316, 234

\bibitem[{{Ricci} {et~al}\mbox{.}(2006){Ricci}, {Prandoni}, {Gruppioni},
  {Sault}, \& {de Zotti}}]{ricci2006}
{Ricci} R., {Prandoni} I., {Gruppioni} C., {Sault} R.~J., {de Zotti} G., 2006,
  \aap, 445, 465

\bibitem[{{Sahayanathan} \& {Godambe}(2012)}]{sahayanathan2012a}
{Sahayanathan} S., {Godambe} S., 2012, \mnras, 419, 1660

\bibitem[{{Sambruna} {et~al}\mbox{.}(1997){Sambruna}, {Urry}, {Maraschi},
  {Ghisellini}, {Mukherjee}, {Pesce}, {Wagner}, {Wehrle}, {Hartman}, {Lin}, \&
  {von Montigny}}]{sambruna1997}
{Sambruna} R.~M. {et~al.}, 1997, \apj, 474, 639

\bibitem[{{Schwartz} {et~al}\mbox{.}(2006){Schwartz}, {Marshall}, {Lovell},
  {Murphy}, {Bicknell}, {Birkinshaw}, {Gelbord}, {Georganopoulos}, {Godfrey},
  {Jauncey}, {Perlman}, \& {Worrall}}]{schwartz2006}
{Schwartz} D.~A. {et~al.}, 2006, \apj, 640, 592

\bibitem[{{Sikora} {et~al}\mbox{.}(2001){Sikora}, {B{\l}a{\.z}ejowski},
  {Begelman}, \& {Moderski}}]{sikora2001}
{Sikora} M., {B{\l}a{\.z}ejowski} M., {Begelman} M.~C., {Moderski} R., 2001,
  \apj, 554, 1

\bibitem[{{Sowards-Emmerd} {et~al}\mbox{.}(2003){Sowards-Emmerd}, {Romani}, \&
  {Michelson}}]{sowards2003}
{Sowards-Emmerd} D., {Romani} R.~W., {Michelson} P.~F., 2003, \apj, 590, 109

\bibitem[{{Sowards-Emmerd} {et~al}\mbox{.}(2004){Sowards-Emmerd}, {Romani},
  {Michelson}, \& {Ulvestad}}]{sowards2004}
{Sowards-Emmerd} D., {Romani} R.~W., {Michelson} P.~F., {Ulvestad} J.~S., 2004,
  \apj, 609, 564

\bibitem[{{Tavecchio} {et~al}\mbox{.}(2002){Tavecchio}, {Maraschi},
  {Ghisellini}, {Celotti}, {Chiappetti}, {Comastri}, {Fossati}, {Grandi},
  {Pian}, {Tagliaferri}, {Treves}, \& {Sambruna}}]{tavecchio2002}
{Tavecchio} F. {et~al.}, 2002, \apj, 575, 137

\bibitem[{{Tingay} {et~al}\mbox{.}(2003){Tingay}, {Jauncey}, {King},
  {Tzioumis}, {Lovell}, \& {Edwards}}]{tingay2003}
{Tingay} S.~J., {Jauncey} D.~L., {King} E.~A., {Tzioumis} A.~K., {Lovell}
  J.~E.~J., {Edwards} P.~G., 2003, \pasj, 55, 351

\bibitem[{{Ueda} {et~al}\mbox{.}(2001){Ueda}, {Ishisaki}, {Takahashi},
  {Makishima}, \& {Ohashi}}]{ueda2001}
{Ueda} Y., {Ishisaki} Y., {Takahashi} T., {Makishima} K., {Ohashi} T., 2001,
  \apjs, 133, 1

\bibitem[{{Urry} \& {Padovani}(1995)}]{urry1995}
{Urry} C.~M., {Padovani} P., 1995, \pasp, 107, 803

\bibitem[{{Villata} {et~al}\mbox{.}(1997){Villata}, {Raiteri}, {Ghisellini},
  {de Francesco}, {Bosio}, {Latini}, {Bucciarelli}, {Chiaberge}, {Chiumiento},
  {Cora}, {Lanteri}, {Lattanzi}, {Massone}, {Peila}, {Racioppi}, {Smart}, \&
  {Scaltriti}}]{villata1997}
{Villata} M. {et~al.}, 1997, \aaps, 121, 119

\bibitem[{{Whiting} {et~al}\mbox{.}(2003){Whiting}, {Majewski}, \&
  {Webster}}]{whiting2003}
{Whiting} M.~T., {Majewski} P., {Webster} R.~L., 2003, \pasa, 20, 196

\bibitem[{{Wiren} {et~al}\mbox{.}(1992){Wiren}, {Valtaoja}, {Terasranta}, \&
  {Kotilainen}}]{wiren1992}
{Wiren} S., {Valtaoja} E., {Terasranta} H., {Kotilainen} J., 1992, \aj, 104,
  1009

\bibitem[{{Wright} {et~al}\mbox{.}(2009){Wright}, {Chen}, {Odegard}, {Bennett},
  {Hill}, {Hinshaw}, {Jarosik}, {Komatsu}, {Nolta}, {Page}, {Spergel},
  {Weiland}, {Wollack}, {Dunkley}, {Gold}, {Halpern}, {Kogut}, {Larson},
  {Limon}, {Meyer}, \& {Tucker}}]{wright2009}
{Wright} E.~L. {et~al.}, 2009, \apjs, 180, 283

\bibitem[{{Zhang} {et~al}\mbox{.}(2010){Zhang}, {Collmar}, {Torres}, {Wang},
  {Lang}, \& {Zhang}}]{zhang2010}
{Zhang} S., {Collmar} W., {Torres} D.~F., {Wang} J.-M., {Lang} M., {Zhang}
  S.-N., 2010, \aap, 514, A69

\bibitem[{{Zhang} {et~al}\mbox{.}(1994){Zhang}, {Marscher}, {Aller}, {Aller},
  {Terasranta}, \& {Valtaoja}}]{zhang1994}
{Zhang} Y.~F., {Marscher} A.~P., {Aller} H.~D., {Aller} M.~F., {Terasranta} H.,
  {Valtaoja} E., 1994, \apj, 432, 91

\end{thebibliography}


\end{document}